# AstroPortal: An ontology repository concept for astronomy, astronautics and other space topics

Robert J. Rovetto

New York, USA
Italy, European Union

rrovetto@terpalum.umd.edu
ontologos@yahoo.com
http://orcid.org/0000-0003-3835-7817
https://purl.org/space-ontology
https://ontospace.wordpress.com
https://ontologforum.org/index.php/RobertRovetto

## Abstract

This paper describes a repository for ontologies of astronomy, astronautics, and other space-related topics. It may be called AstroPortal (or SpacePortal), AstroHub (or SpaceHub), etc. The creation of this repository will be applicable to academic, research and other data-intensive sectors. It is relevant for space sciences (including astronomy), Earth science, and astronautics (spaceflight), among other data-intensive disciplines. The repository should provide a centralized platform to search, review and create ontologies for astro-related topics. It thereby can decrease research time, while also providing a user-friendly means to study and compare knowledge organization systems or semantic resources of the target domains. With no apparent repository available on the target domain, this paper also expresses a novel concept.

## 1. Introduction

This paper conceptualizes a repository for ontologies of astronomy, astronautics, and other space-related topics. It may be called AstroPortal (or SpacePortal), AstroHub (or SpaceHub), etc. Although a focus for this paper is ontologies, the repository concept may be for any other so-called knowledge organization system, e.g., controlled vocabularies, classifications, thesauri, taxonomies, knowledge graphs, etc. Each have their own, sometimes overlapping, features, functionalities and purposes.

Astronomy, astronautics, and other outer space disciplines collect and produce large volumes of data. The data, domain knowledge, and the overall topics can be modeled in various ways, from distinct perspectives, and degrees of abstraction. This applies to any discipline or topic area being modeled.

See also: https://github.com/rrovetto/astroportal

Ontology development is one means to represent or model concepts, knowledge, beliefs, facts, statements, meaning and semantics. It is applied to databases, information systems, and other data-intensive disciplines. Computational ontologies and so-called 'knowledge graphs', are produced in various ways and degrees of complexity for artificial intelligence, semantic web, natural language processing, model-based systems engineering, and other research areas.

Some existing repositories [1] [2] or software for cataloging some models include OntoPortal and its the domain-specific repositories [3], OntoHub [4], and OntologyDesignPatterns.org [5]. All of these repositories can be generic or topic-specific. The OntoPortal software platform (2019), for example, is a software package that can be used to create web-based repositories.

Some examples of ontologies and vocabularies for space applications and topics are described in [6], and include the authors space domain ontologies [6-9], and the Ontology of Astronomical Object Types [5]. However, compared to other disciplines, there appear to be significantly less knowledge models for this topic, at least publicly accessible. NASA and other agencies, for instance, have a history of creating ontologies or similar knowledge representation systems for use in autonomous vehicles, or model-based systems engineering projects, etc. Among the author's project concepts for space domain knowledge modeling, is a working catalog of existing knowledge organization systems [11][12]. By contrast to the astronomy and astronautics scope of this concept, a project by a similar name (SpacePortal [13]) was proposed for spatial cognition. However, my broader modeling interests include concepts of space (generally construed), spatiality and spatio-temporality. Concepts for or relevant to an AstroPortal were also described in [18-20]

The following sections summarize the purpose, features and principles, as well as desired capabilities for the AstroPortal repository concept.

For the purposes of this communication, 'ontologies' may be understood broadly to include any knowledge organization system (KOS), including taxonomies, classification schemes, computational ontologies, thesauri, knowledge models, knowledge graphs, conceptual models, and semantic models.

## 2. Purpose

The purpose of AstroPortal is to store, and provide access to, distinct conceptual, linguistic, semantic, ontological, and other knowledge models or knowledge organization systems (KOS) for astro- or related space domains. These are broadly construed to non-exhaustively include astronomy, astronautics, space policy and law, and their sub-disciplines or branches. Persons, projects, groups, and organizations interested in creating and curating these models can develop and add them to the repository.

## 3. Relevance

Astronomy, astronautics, and other related topics have data and generic information of various sorts. This is one area of relevance for the knowledge models. Types of data/information include

By Robert J. Rovetto – rrovetto@terpalum.umd.edu , ontologos@yahoo.com

See also: https://github.com/rrovetto/astroportal

observational data from telescopes and other devices; telemetry data of spacecraft and other devices; space systems engineering and requirements information; solar system information; generic knowledge about the respective topic or discipline; predictive and simulation data; qualitative aspects that can be represented; etc. In other words, AstroPortal is relevant for academic, research and other data-intensive sectors; as well as activities and disciplinary areas such as space sciences (including astronomy), Earth science, and astronautics (spaceflight).

## 4. Accessibility, Scope and Specificity in Design

A number of aspects to consider for dynamic repositories or portals for astro-semantic systems, i.e., KOS, are display in the following figure. The degree of accessibility of an AstroPortal repository important for its development, including any potential financial costs for that access. For example, it may be publicly- and freely accessible, or publicly-accessible for a fee. Alternatively it may be privately or internally accessible.

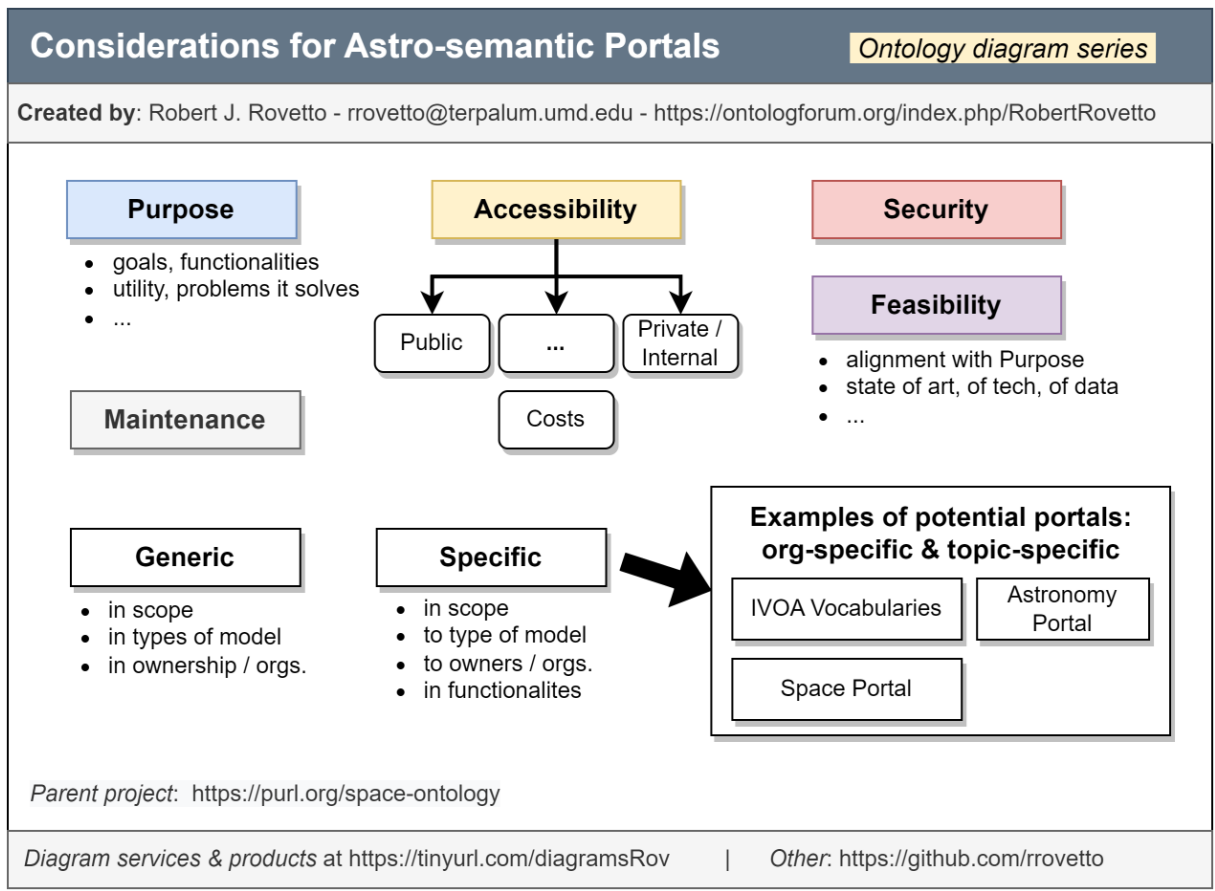

Considerations of generality or specificity are important as well. AstroPortals may be generic with respect to scope, to the type of KOS, to ownership of a given (typeo of) KOS, etc. Alternatively, it may be specific to a particular domain, type of KOS (e.g., only ontologies), or to a given project or owner (e.g., to the IVOA).

By Robert J. Rovetto – rrovetto@terpalum.umd.edu , ontologos@yahoo.com

See also: https://github.com/rrovetto/astroportal

For a generic (or broad) scope, an AstroPortal may include various outer space topics. Alternatively, it may be specific to a particular discipline, such as Astronomy. The name 'AstroPortal' may or may not scope astronomy, specifically. Similarly, 'SpacePortal' may reflect a broader scope, including other outerspace topics and semantic artifacts.

## 5. Features / Guidelines

Some desired features, subject to revision, for AstroPortal are as follows.

**Neutrality**. AstroPortal should be both neutral with respect to tools, computable languages and to the models (ontologies, taxonomies, etc.) themselves, including any assumptions they make. It should not favor or bias a particular ontology development environment, methodology, ontology, semantic model, or set of ontological commitments. It should not assume or bias a particular formal semantics; or a semantic, philosophical, ideological or other starting assumption.

**Wide scope.** AstroPortal should have a relatively wide scope, including models from various astro-related topics and disciplines. The scope can be narrower, however, by forming distinct repositories for each astro-specific topic.

*Note*: Scoping is a challenge unto itself because there will be overlap with other widely applicable disciplines (e.g., physics). An ontology of interstellar particle phenomena, for example, will likely overlap in topic or content, with supposed ontologies of say fluid mechanics and chemistry. If the latter are applied to or about astro data, then it would be within the repository scope.

**Diversity of Modeling**. AstroPortal explicitly acknowledges the fact that *more than one conceptualization and model is possible for the same topic*, i.e., distinct ontologies, classification schemes, taxonomies, can be developed (each with potentially unique or contrasting starting assumptions and commitments). In other words, AstroPortal acknowledges the fact that this cliché 'Don't reinvent the wheel' is not mandatory (and arguably not applicable) for ontologies (or ontology use), in part, because the preceding sentence, but also for ethical responsibility (e.g., prevention of monopolizing attitudes and actions by any single model or group; and to preserve creativity, freedom, and innovation).

## 6. Functional Capabilities

The following are a non-exhaustive set of capabilities AstroPortal should be able to perform.

AstroPortal should be capable of running locally on one's own computer, as an application storing distinct ontologies, displaying their features, statistics, and other functionalities. This is useful for ontology owners who are unable to, or who do not wish to, make their ontology public. AstroPortal should also have a public-facing web-based platform for the same with the ability to make private (i.e. not public-facing) any added or published ontologies. Other capabilities include the following.

By Robert J. Rovetto – rrovetto@terpalum.umd.edu , ontologos@yahoo.com

See also: https://github.com/rrovetto/astroportal

**Ontology creation** – developers may be able to create their own ontology(s). An interesting potential capability may be automatic term-matching. That is, when a user manually enters a new category term, the system can compare the string being entered with any identical or similar terms and dynamically display them. This should be an optional feature the user can toggle on and off, but will allow them to compare other terms, potentially identifying differences in conceptualization, methodology, ideological assumptions, definitions, higher-level inherent abstract characterizations (e.g., from a more abstract ontology), ontological commitments, etc.

Note: this is arguably the least central functionality because as a repository, AstroPortal should be primarily about viewing ontologies.

**Upload ontologies** – users can upload an ontology to the repository in various ontology file formats

**Browse ontologies** – users can browse a listing of ontologies. Ontologies that are set as private should not be displayed publicly.

**Search for ontologies** – users can search for specific ontologies by entering a phrase in a search box.

**Search for ontology elements/constructs (e.g., category terms)** – users can search for specific ontology elements or constructs, such as ontology classes/categories, relations, etc., by entering a phrase in a search box.

**Supports many computable languages** – in accord with the language-agnostic guideline, AstroPortal should be able to store and process ontologies that are formalized in various artificial languages (e.g., RDF[14], OWL[15], CLIF[16], KIF[17], ORM, DOL, CASL, etc.) Non-ontology languages (e.g., markup, script, or data structure languages as XML, JSON, etc.) may also be desired.

**Compare ontologies and their elements** – examine the similarities and differences between ontologies or between elements from distinct ontologies. This includes comparing ontology terms themselves (phrasing, labels), definitions (of various sorts), assumptions and commitments (e.g., any imported ontologies at equal, less or greater degree of abstraction, etc.).

Some functions of repositories such as OntoPortal, as presented on the internet, provide Browse / Search, Annotate, Publish, Mappings, etc. Ontohub repository allows for various formalisms and ontology file formats.

---

In short, if the above-mentioned functions and capabilities are coupled with user-friend graphical interface that is accessible in any geographic region, then AstroPortal (or SpacePortal) should provide value to anyone interested in developing or understanding ontologies and other knowledge organization systems for space-related topics.

By Robert J. Rovetto – rrovetto@terpalum.umd.edu , ontologos@yahoo.com

See also: https://github.com/rrovetto/astroportal

## 7. Closing

This paper has introduced a concept for one or more repositories to store (at least display) and interact with astronomy, astronautics and other space-related terminological, knowledge modeling organization systems such as computational ontologies and taxonomies. Depending on the context as well as scoping and other considerations, one or more dynamic repositories (or portals) are feasible. Some desired features and capabilities of AstroPortal (or SpacePortal) were summarized. As a preliminary concept paper, future work may include more detailed research, and refinement of requirements, feasibility and functionalities. As currently unfunded work, formal work collaborations or sponsors are welcome for the development of such repositories and to explore its potential utility for space-related disciplines.

By Robert J. Rovetto – rrovetto@terpalum.umd.edu , ontologos@yahoo.com

See also: https://github.com/rrovetto/astroportal

By Robert J. Rovetto – rrovetto@terpalum.umd.edu , ontologos@yahoo.com